\documentclass{article}
\usepackage[T1]{fontenc}
\usepackage{graphicx}
\usepackage{caption}
\usepackage{xurl} 
\usepackage{geometry}
\usepackage[style=numeric,minbibnames=3,maxbibnames=3,sorting=none,backend=bibtex]{biblatex}
\usepackage{amsmath} 
\usepackage{geometry}
\usepackage{gensymb}
\usepackage{subcaption}
\usepackage{authblk}

\title{Performance and uses of the Refimeve metrological signal, 1000 km from the source.}
\author[1]{Mathieu Collombon}
\author[2]{Etienne Cantin}
\author[1]{Gaëtan Hagel}
\author[3]{Paul Eric Pottie}
\author[1]{Marie Houssin}
\author[1]{Caroline Champenois}

\affil[1]{Laboratoire PIIM, Aix-Marseille Université, CNRS, Marseille, France}
\affil[2]{Laboratoire de Physique des Lasers (LPL), Université Sorbonne Paris Nord, CNRS, Villetaneuse, France}
\affil[3]{Laboratoire Temps Espace (LTE), Observatoire de Paris, Université PSL, Sorbonne Université, Université de Lille, LNE, CNRS, Paris, France}

\begin{document}
\maketitle

\begin{abstract} 
We present a detailed analysis of the end-user performance of the metrological signal at 1542 nm disseminated by the French national fibre network Refimeve, about 1000 km from the source. By the mean of a local ultrastable laser at 729 nm and a frequency comb, we are able to carry out stability and phase noise measurements of the signal with respect to the local laser. With a focus on phase noise analysis we identify different timescales of interest for the use of this signal in optical frequency metrology.
\end{abstract}
\section{Introduction}
Since its first development in the 90's \cite{ma_delivering_1994}, the technique to correct for the optical phase noise induced by mechanical stress in fibres has been an essential tool for the progress of frequency metrology, rapidly leading to phase-coherent optical frequency disseminations \cite{newbury_coherent_2007}. 
Since a few years the performances of these links have reached a level enabling remote comparison of state-of-the-art atomic clocks \cite{lisdat_clock_2016,guena_first_2017} hence opening new opportunities for the future of time transfer and time keeping\cite{dimarcq_roadmap_2024}. These coherent remote comparisons are also providing greater precision in fields such as fundamental physics \cite{delva_test_2017}, geodesy \cite{takano_geopotential_2016,lisdat_clock_2016,grotti_geodesy_2018} or astronomy \cite{clivati_common_clock_2020}. Furthermore, these stabilized links allow for the dissemination of ultrastable light by National Metrology Institutes (NMIs), enabling distant users to have a direct access to SI-traceable, ultrastable and accurate optical signals without the expense of a dedicated experimental setups. This has permitted great results in the field of molecular \cite{santagata_high-precision_2019-1,votava_comb_2022} and atomic \cite{friebe_remote_2011,matveev_precision_2013,morzynski_absolute_2015} spectroscopy, showing that connecting NMIs to other national (and international) laboratories strongly enhance the capabilities of the spectroscopy community.\\  
The Refimeve research infrastructure \cite{refi_web} is a French initiative that participates in shaping this scientific landscape in France and Europe. Started in 2012 by the NMI Time and Space Laboratory (LTE), and the Laser Physics Lab (LPL), the project aimed at the dissemination of an ultrastable, absolute optical reference at 1542~nm. This was made possible by the development of active phase noise compensation directly on internet fibres without any traffic disturbances \cite{lopez_cascaded_2010,guillou-camargo_first_2018}, making the deployment of a national-wide network with a multi-access structure far more easy \cite{bercy_ultrastable_2016}. The Refimeve network is using the optical fibers of RENATER, the French National Telecommunications Network for Technology, Education and Research. As of today about thirty laboratories are part of the network and receive the so-called Refimeve signal, an optical carrier referenced to the SI-frequency standard with a residual drift of less than 10~mHz$\cdot$s$^{-1}$ and a fractional uncertainty that can reach $10^{-19}$ \cite{cantin_accurate_2021}. Neighbors NMIs are also connected to the network (Italy's INRIM, Germany's PTB and UK's NPL) enabling fruitful collaborations and international measurement campaign \cite{clivati_coherent_2022,schioppo_comparing_2022}.\\
Our lab in Marseille, south of France, is hosting several experimental setups based on Calcium spectroscopy. One of these experiments is focusing on high precision laser spectroscopy of Ca$^+$ and takes advantage of the long-lived D$_{5/2}$ state, with a lifetime of 1.165~s \cite{shao_precision_2016}. To exploit the clock transition at 729~nm \cite{knoop_metastable_2004} we've developed a Local Ultra Stable laser (LUS). It is used both for one-photon excitation of the clock transition and in a three-photon coherent excitation process referenced to the magnetic dipolar transition D$_{3/2} \to $ D$_{5/2}$ \cite{champenois_terahertz_2007,collombon_experimental_2019}.\\
It is in this context that we became partner of the Refimeve infrastructure, receiving the Refimeve signal approximately 1000 km from its source in Paris (LTE). The link is segmented in two parts (Paris-Lyon, 600 km and Lyon-Marseille, 400 km) in order to mitigate the fundamental limitations of phase noise cancellation in fibre \cite{newbury_coherent_2007}, where the loop bandwidth is inversely proportional to the length of the link. In this paper we show an analysis of the signal's quality that we are able to investigate using a local ultrastable laser. After a description of the experimental setup we show detailed measurements of the comparison between our local laser and the Refimeve signal, in terms of fractional frequency stability, phase noise and laser linewidth. Throughout this analysis we are able to identify three relevant timescales where the Refimeve signal can be of different uses for a distant user.\\
 
\section{Experimental setup}
We routinely use a laser at 729 nm locked on a high finesse Fabry-Pérot cavity, in combination with an optical frequency comb (OFC), to perform coherent spectroscopy on Calcium ions \cite{collombon_experimental_2019}. Here we quickly describe this laser and the method used to measure its stability either using local resources or using the Refimeve signal and OFC.
\begin{figure}[h!]
\centering\includegraphics[width=1\textwidth]{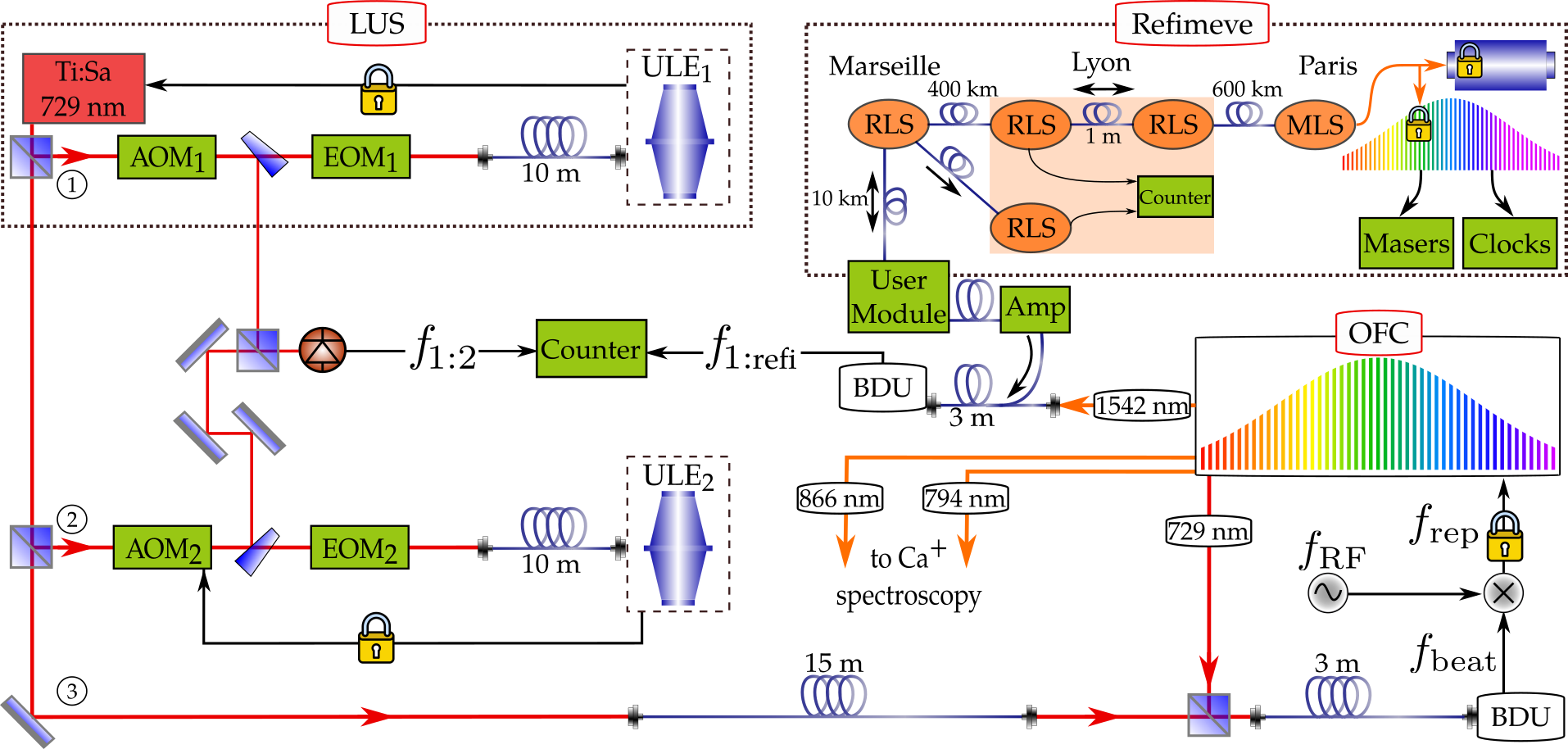}
\captionsetup{width=0.9\textwidth}
\caption{Schematic view of the experimental setup. LUS: Local UltraStable laser, AOM : Acousto-Optic Modulator, EOM : Electro Optic Modulator, ULE: Ultra Low Expansion glass Fabry Pérot cavity, RLS: Repeater Laser Station, MLS: Multibranch Laser Station, AMP: Lumibird amplifier, Counter: Multichannel synchronous phase counter, OFC: Optical Frequency Comb, BDU: Beat Detection Unit, $2\omega$: frequency doubling.}
\label{fig:setup}
\end{figure}
Figure \ref{fig:setup} shows a schematic view of the experimental setup. The beam of a homemade Ti:Sa laser emitting at 729 nm is split in three paths. The first one is frequency shifted by an Acousto Optic Modulator (AOM$_1$, $f_1$=110 MHz) and phase modulated at 40 MHz with an Electro Optic Modulator (EOM$_1$). After traveling in a 10~m uncompensated fibre to a separated room, the light is coupled to the TEM$_{00}$ mode of a high finesse 15~cm long Fabry-Pérot cavity, ULE$_1$ (Finesse F$_1$ $\approx$165000, and resonance linewidth $\Delta\nu\approx$ 6 kHz). The vacuum chamber hosting the cavity is enclosed in a temperature controlled ($\pm$ 1 mK) 2 cm thick aluminium box siting on a vibration isolated platform. Fast detection of the reflected light and its subsequent demodulation with the 40~MHz signal fed to EOM$_{1}$ produces the so-called Pound-Drever-Hall (PDH) signal \cite{drever_laser_1983}. This error signal is processed by an analog loop filter (Toptica FALC110) and slow (fast) corrections are applied on the intra-cavity piezo transducer (EOM) with a bandwidth of 200 kHz. In the following we refer to this system as the Local UltraStable laser (LUS).\\   
To have an estimate of its precision a similar setup is used in parallel on a second beam path, with AOM$_2$ ($f_2$=468~MHz) and EOM$_2$ (40 MHz), to couple the light to a second cavity, ULE$_2$. The TEM$_{00}$ mode of this cavity shows a lower photon lifetime leading to a finesse F$_2$ $\approx$90000 and $\Delta\nu\approx$ 11 kHz. This is most likely due to pollution on the mirrors during the recent relocation of the lab, and could not be brought back to its original value of 210000 despite several cleaning. Note that the vacuum chamber hosting this cavity is not enclosed in an aluminium box, nor temperature stabilized, due to the aforementioned cleaning process. The PDH signal generated with this cavity is used for frequency correction on AOM$_2$ with a bandwidth of 200 kHz.\\
By sampling about 4$\%$ of the light at the output of AOM$_1$ and AOM$_2$ and with a proper spatial overlap of the beams we detect a beatnote, $f_{1:2} = f_{2}-f_{1}$. Because the laser fields of these two paths are locked onto independent cavities we assume that, within the loop bandwidth, the electromagnetic fields are uncorrelated once transmitted by the AOMs. Therefore, an analysis of $f_{1:2}$ will show the convolved noise of both cavity/locking loop subsystem. This method gives an upper limit on the performance of the LUS, but hardly allow access to the individual performances of each systems. We further discuss this typical drawback in the next section. In our specific case, with F$_2$ being almost a factor of two lower than F$_1$, we can fairly assume that the first lock based on ULE$_1$ will give the best results.\\

The third beam path, consisting of a 15 m long uncompensated fibre, links the LUS to a commercial frequency comb (TOPTICA-DFG) for stabilization of the OFC to the LUS. This Er:doped fibered laser, with a repetition frequency $f_{\text{rep}}$=80 MHz, relies on difference frequency generation to cancel out the offset frequency $f_{\text{ceo}}$ \cite{fehrenbacher_free_running_2015}. Interference between the LUS and the relevant comb's mode at 729 nm is achieved in a 3 m long fibre for optimal spatial overlap. The subsequent beatnote $f_{\text{beat}}$ is detected with a commercial "beat detection unit" (BDU) consisting of a grating and a fast photodiode.  For technical reason we tune $f_{\text{rep}}$ such that $f_{\text{beat}}$ falls close to 48 MHz. To phase lock $f_{\text{rep}}$ to the LUS we detect the phase difference between $f_{\text{beat}}$ and $f_{\text{RF}}$ from a commercial, GPS referenced synthesizer (HP 8657A), and use it as the error signal to drive an analog loop filter (FALC110). Fast corrections are applied to an intracavity EOM while slow feedback drives the pump current of the fs-oscillator, ensuring a loop bandwidth of about 700 kHz.
When the repetition frequency of the OFC is locked, the frequency stability and phase noise properties of the LUS are transferred to the OFC's entire spectrum \cite{collombon_phase_2019}. We use the \mbox{1542 nm} output of the comb to make a beatnote with the Refimeve signal. The optical frequency reference is transferred from LTE in Paris to PIIM in Marseille through 1000 km of fiber. The fiber noise is actively compensated using a cascaded approach, where compensation is performed segment by segment to ensure sufficient correction bandwidth and signal-to-noise ratio, thanks to regeneration laser stations RLS and multibranch regeneration laser stations MLS \cite{lopez_cascaded_2010,lopez_ultrastable_2012,bercy_ultrastable_2016,cantin_accurate_2021}. The signal is transferred back over each segment to ensure continuous characterization of the transfer. This characterization enables a detailed assessment of the transfer performance, as described in the following sections. The signal arriving in the lab is directly connected to a so called user module in which part of the light is frequency shifted and sent back, ensuring phase noise cancellation up to this point. The output of the user module being less than 100 $\mu W$, an optical amplifier is needed (Lumibird), bringing the power to 5 mW. This light is combined with the comb's light in the same way as previously described for the 729 nm line. In this case the built-in grating of the BDU, with a bandwidth of 10~GHz, is not only useful to limit the number of non-useful comb's modes but also rejects most of the amplified spontaneous emission noise added by the amplifier.\\ 
The beatnote detected on the photodiode, referred as $f_{1:\text{refi}}$ in the following, falls around 26 MHz, depending on the exact frequency of the LUS to which the comb is referenced. Analysis of $f_{1:\text{refi}}$ will give a comparison between our local LUS and the Refimeve source laser, through  1000 km of stabilized fibre and a frequency comb.\\

Both beatnotes, $f_{1:2}$ and $f_{1:\text{refi}}$ are sent to a K+K phase counter \cite{kramer_multi_channel_2001} working in dead-time free mode ($\Lambda$ gate). In the following sections we present a detailed analysis of these quantities, both in terms of frequency stability and phase noise.

\section{Frequency stability}
\begin{figure}[ht!]
\centering\includegraphics[width=1\textwidth]{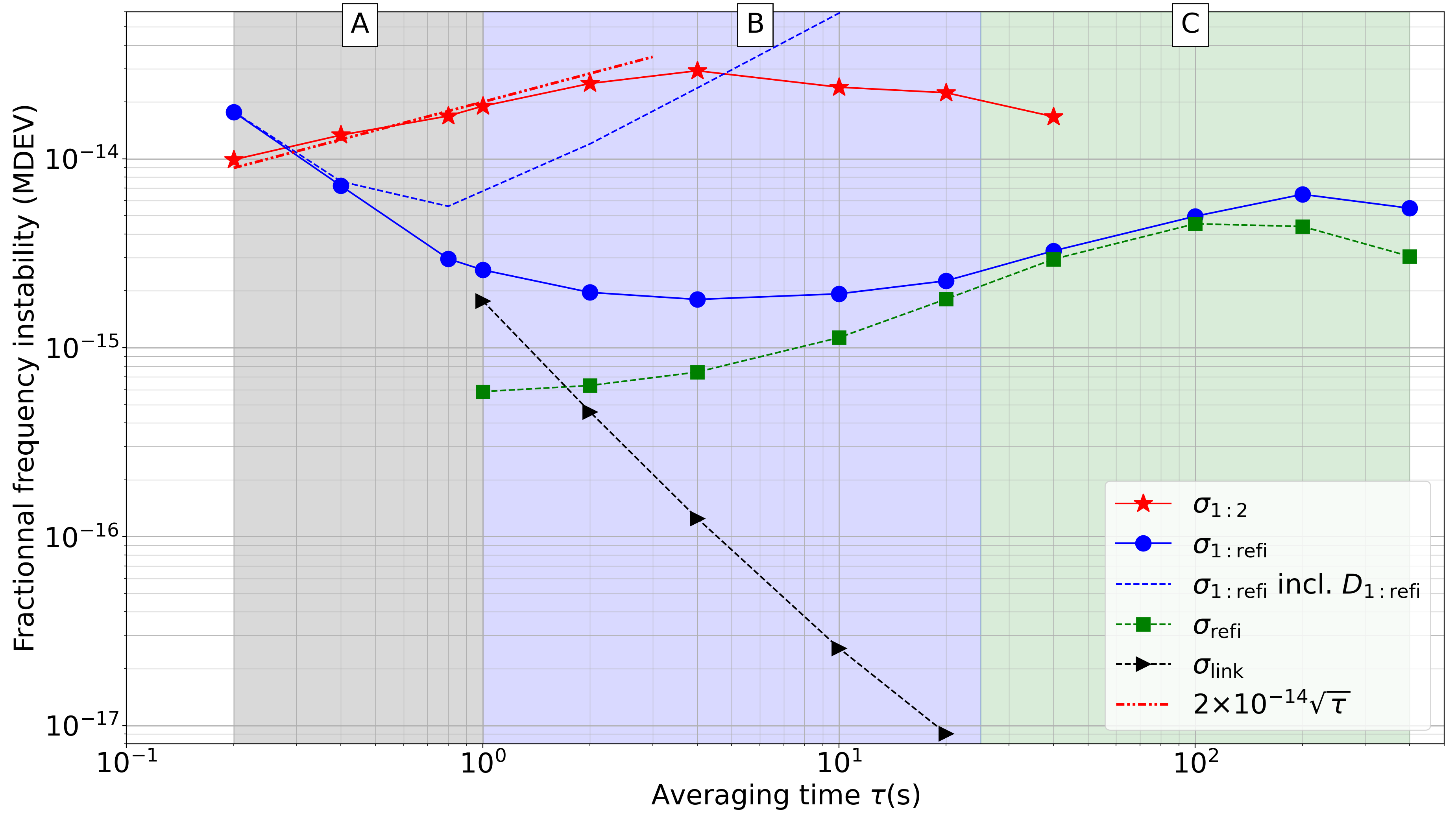}
\captionsetup{width=0.9\textwidth}
\caption{Fractional frequency instability $\sigma$ expressed with the Modified Allan deviation (MDEV) vs averaging time $\tau$. Blue dashed line:  $\sigma_{1:\text{refi}}$ from $f_{1:\text{refi}}$ including $D_{1:\text{refi}}$; blue circles: $\sigma_{1:\text{refi}}$ from $f_{1:\text{refi}}$ corrected for $D_{1:\text{refi}}$; red stars: $\sigma_{1:2}$ from $f_{1:2}$ corrected for $D_{1:2}$; green square: $\sigma_{\text{refi}}$ from the Refimeve source laser; black triangles: $\sigma_{\text{link}}$ from the fibre link instability.}
\label{fig:stab}
\end{figure}

The frequency stability of the LUS is estimated locally by measuring the frequency $f_{1:2}$ with a gate time $\tau_0 = 200$ ms. The raw data are showing a significant frequency drift of $D_{1:2}=-13.6$ Hz$\cdot$s$^{-1}$. This drift corresponds to the relative thermal behaviour of both cavities, however most of it can be attributed to ULE$_2$ as this cavity does not benefit of the same thermal shielding/control as its counterpart. This drift is removed before calculating the modified Allan deviation (MDEV), $\sigma_{1:2}$ displayed on fig.\ref{fig:stab} (red stars). We can observe a rising slope behaving like $\approx2\times10^{-14}\sqrt{\tau}$, between 0.2 s and 1 s, corresponding to a random walk frequency. This phenomenon, in combination with the poor finesse of ULE$_2$, are setting an upper bound on the estimation of the LUS fractional frequency instability, between $1\times10^{-14}$ and $3\times10^{-14}$. We are confronted here to the drawback of this two-cavities method to estimate the performances of a stabilized laser : this measurement alone cannot distinguish the individual contributions of ULE$_1$ and ULE$_2$. 
However with the use of the Refimeve signal we can circumvent this problem and push further the diagnosis of the LUS. We record the frequency $f_{1:\text{refi}}$ with the same gate time $\tau_0$ and observe a linear frequency drift $D_{1:\text{refi}}=1.58$ Hz$\cdot$s$^{-1}$. With a comb scaling factor of 2.11 between 1542~nm and 729~nm this corresponds to 3.33 Hz$\cdot$s$^{-1}$ at 729~nm. Given the very low residual frequency fluctuations of the source in LTE \mbox{(< 10 mHz$\cdot$s$^{-1}$)} \cite{rodolphe_private} we can attribute this measured drift solely to the thermal expansion of ULE$_1$. 
The blue circles data points on fig.\ref{fig:stab} shows the MDEV $\sigma_{1:\text{refi}}$ with $D_{1:\text{refi}}$ removed before calculation while the dashed blue line shows the corresponding MDEV with $D_{1:\text{refi}}$ . Comparing these two results shows that $\sigma_{1:\text{refi}}$ is about 10 times lower than $\sigma_{1:2}$ at 3 s, confirming that ULE$_2$ is limiting in the local diagnosis of the LUS. We can get further insight on $\sigma_{1:\text{refi}}$ by plotting the stability of the Refimeve laser source and the fibre link noise, respectively green square and black triangle on fig. \ref{fig:stab}. This allows us to identify three regimes.\\ Regarding the long-term behaviour for averaging time $12$ s$<\tau<300$ s and labeled C on fig. \ref{fig:stab} (green overlay), we can see that $\sigma_{1:\text{refi}}$ fairly reproduces $\sigma_{\text{refi}}$. This shows that in this time span the Refimeve signal provides long-term stability useful to measure and remove the linear drift. Once drift-corrected the LUS stability is not limiting and does not show any particular noise behaviour such as random walk frequency.\\ On mid-timescale corresponding to $1$ s$<\tau<12$ s (B label and blue overlay on fig. \ref{fig:stab}), $\sigma_{1:\text{refi}}$ reaches a flicker frequency plateau at $2\times10^{-15}$. At this level neither the Refimeve source laser nor the fibre link should be limiting, therefore showing that this plateau is the ultimate stability that can be reached by the LUS and transferred to the OFC. 
Finally we identify a third region on short timescale: $200$ ms$<\tau<3$s (label A and grey overlay on fig. \ref{fig:stab}). In this regime the comparison between $\sigma_{1:\text{refi}}$ and the extrapolated $\sigma_{\text{link}}$ indicates that the fibre link most likely dominates the instability and that the LUS could be of use to actually characterize the link performances. To clarify the dynamics on these short timescales phase noise analysis is more adapted and is presented in the following section.

\section{Phase analysis}
\subsection{Noise spectrum}
We present here the analysis of the same beatnotes, $f_{1:2}$ and $f_{1:\text{refi}}$, but we focus on the phase noise of these signals, expressed as the power spectral density (PSD) $S_{\varphi}(f)$ of their residual phase jitters $\varphi(t)$.
\begin{figure}[h!]
\centering\includegraphics[width=1\textwidth]{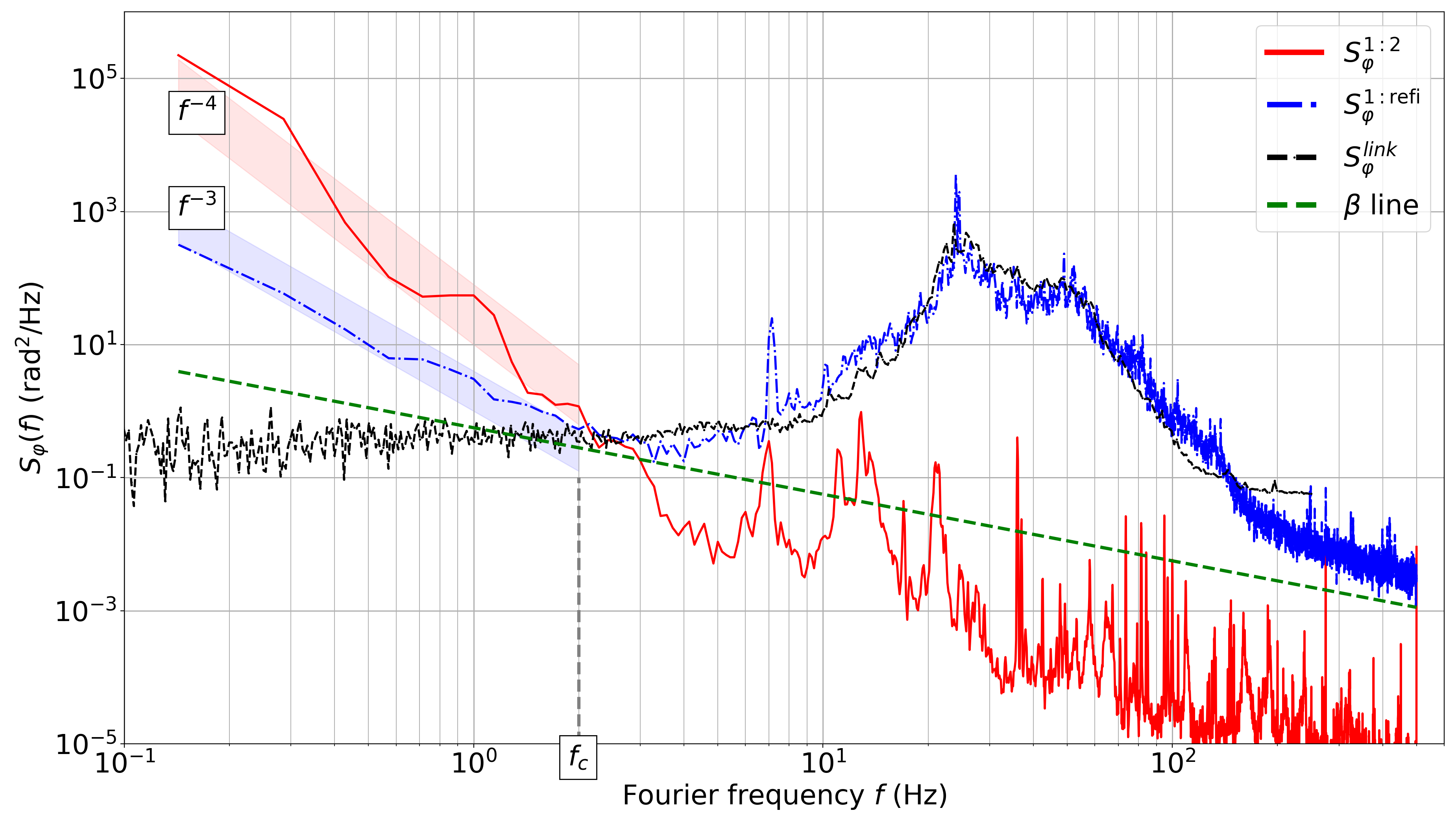}
\captionsetup{width=0.9\textwidth}
\caption{Power Spectral Density $S_{\phi}(f)$ vs Fourier frequency. Red, plain line : $S_{\phi}^{1:2}(f)$ from $f_{1:2}$;  blue dashed-dotted line: $S_{\phi}^{1:\text{refi}}(f)$ from $f_{1:\text{refi}}$; black dashed line: $S_{\phi}^{\text{link}}(f)$; green dashed line: $\beta$ separation line, see text for equation.}
\label{fig:phase_noise}
\end{figure}
The phase counter is set to its shortest gate-time, $\tau_0=1$ ms, and fourteen consecutive phase sets of 7 s each are processes and analyzed. The raw data show a typical linear slope of $\omega t$ with $\omega$ the angular frequency of the signals. Once this running phase is removed we can see a quadratic behavior of the phase record. This corresponds to a linear frequency drift and can be express as $\frac{1}{2}\times D t^2$, with $D$ the drift in Hz$\cdot$s$^{-1}$. This component is fitted and removed before  calculation of $S_{\varphi}(f)$ for each sets. After averaging we get $S_{\varphi}^{1:2}$ from $f_{1:2}$ and $S_{\varphi}^{1:\text{refi}}$ from $f_{1:\text{refi}}$, respectively in red solid line and in blue dash-dot line on figure \ref{fig:phase_noise}. We also show on the same plot the phase noise from the fibre link, $S_{\varphi}^{\text{link}}$, and the beta-separation line, $\beta(f)$ = $\frac{8\text{ln}(2)}{f\pi^2}$ \cite{di_domenico_simple_2010}. This line is a powerful tool to asses the magnitude of the phase noise as when $S_{\varphi}(f)$>>$\beta(f)$ the phase excursion are greater than one cycle.\\ We can identify a corner frequency $f_c=2$ Hz below which $S_{\varphi}^{1:2}$ shows a significant noise with a strong rising slope in $f^{-4}$, signature of a random walk frequency and consistent with the fractional frequency previously presented. White frequency noise ($f^{-2}$) dominates the power spectral density for Fourier frequencies higher than $f_c$, eventually leading to $S_{\varphi}^{1:2} < \beta$.
Regarding $S_{\varphi}^{1:\text{refi}}$, it is dominated by flicker frequency noise ($f^{-3}$) for frequencies lower than $f_c$. Given the noise limit $S_{\varphi}^{link}$ set by the link, we can conclude that this noise comes from our user end. It is most likely arising from the RF amplification processes before phase counting, but we cannot rule out a noise coming from the LUS. For frequencies larger than $f_c$ we observe a quick rising of the noise, forming a bump between 20 Hz and 70 Hz. This is a typical signature marking the cutoff frequency from the servo loops for the active phase noise compensation of the link. Note that we get the noise from the link from the Refimeve monitoring system where it's obtained from an out of loop measurement, therefore sampling twice the noise of the link \cite{Koke_2019}. $S_{\varphi}^{\text{link}}$ is rescaled accordingly to represent the noise of one fiber.  
We can notice that, for frequencies higher than $f_c$, $S_{\varphi}^{1:\text{refi}}$ reproduces $S_{\varphi}^{\text{link}}$, showing again that the LUS can be of use to probe the Refimeve signal and asses the quality of the noise compensation directly on the user side. At Fourier frequencies lower than  $f_c$ we are using the Refimeve signal as a reference to monitor the performances of the LUS. 

\subsection{Linewidth}
\begin{figure}[ht!]
\centering\includegraphics[width=1\textwidth]{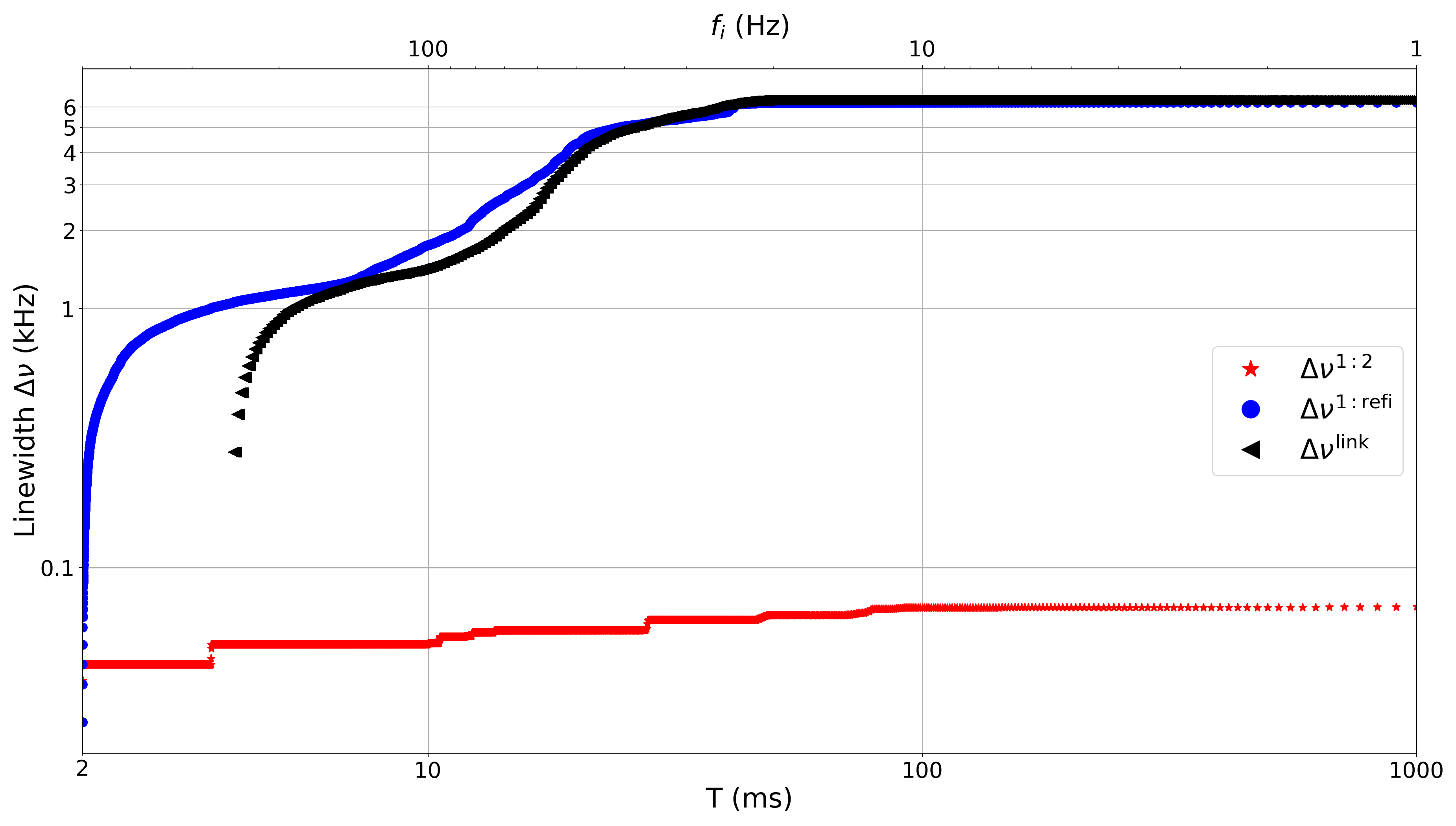}
\captionsetup{width=0.9\textwidth}
\caption{Computed linewidth $\Delta \nu$ according to Eq. (\ref{eqn:linewidth}) as function of the integration time $T$ (bottom axis) or the Fourier frequency $f_i$ (top axis). Red stars: $\Delta \nu^{1:2}$ from $S_{\varphi}^{1:2}$, blue circles: $\Delta \nu^{1:\text{refi}}$ from $S_{\varphi}^{1:\text{refi}}$ and black triangle $\Delta \nu^{\text{link}}$ from $S_{\varphi}^{\text{link}}$.}
\label{fig:linewidth}
\end{figure}

Thanks to the Beta-separation line formalism introduced earlier one can also access the linewidth of a laser directly from its phase noise measurement. When $S_{\varphi}(f)$ is greater than $\beta(f)$ the noise effectively broadens the laser \cite{di_domenico_simple_2010}. Following this we can evaluate the linewidth $\Delta \nu$ for an integration time $T$ with :
\begin{equation}
\label{eqn:linewidth}
\Delta\nu(T) = \sqrt{8\text{ln}(2)A(T)}
\end{equation}
with
\begin{equation}
\label{eqn:integral}
A(T) =  \int_{f_i=1/T}^{f_n=500} S_{\varphi}(f) H[S_{\varphi}(f)-\beta(f)] f^2 \,df  
\end{equation}
where H denote the Heaviside function.\\
Here we follow the method used in \cite{tonnes_PhD} and compute the integral for different lower bound frequency $f_i=1/T$ while the upper bound frequency f$_n$= 500 Hz is kept constant and equal to the Nyquist frequency set by the counter's gate time $\tau_0=1$ ms. This allow us to associate an integration time $T$ to the computed linewidth $\Delta \nu$ in order to identify the timescale of the broadening processes hence determine how long it takes for the laser to reach its stationary linewidth.\\ Figure \ref{fig:linewidth} displays the evolution of the computed linewidth as function of $T$. The results are showing that the received  Refimeve optical carrier reaches a maximum linewidth of about \mbox{6.2 kHz} after 40 ms of integration. Both $\Delta \nu ^{1:\text{refi}}$ and $\Delta \nu^{\text{link}}$ are in very good agreement for integration time longer than 25 ms (40 Hz), which is consistent with the typical timescale of the servo loop for the link phase noise compensation. Regarding $\Delta \nu ^{1:2}$ it represents the convolved linewidth of the two laser path phase locked on each cavities. It stays below 80 Hz linewidth in a 1 s integration window but we cannot push further the estimation due to a difference between the two cavities mentioned before.

\section{Local uses and perspectives}
\begin{figure}[ht!]
\centering\includegraphics[width=1\textwidth]{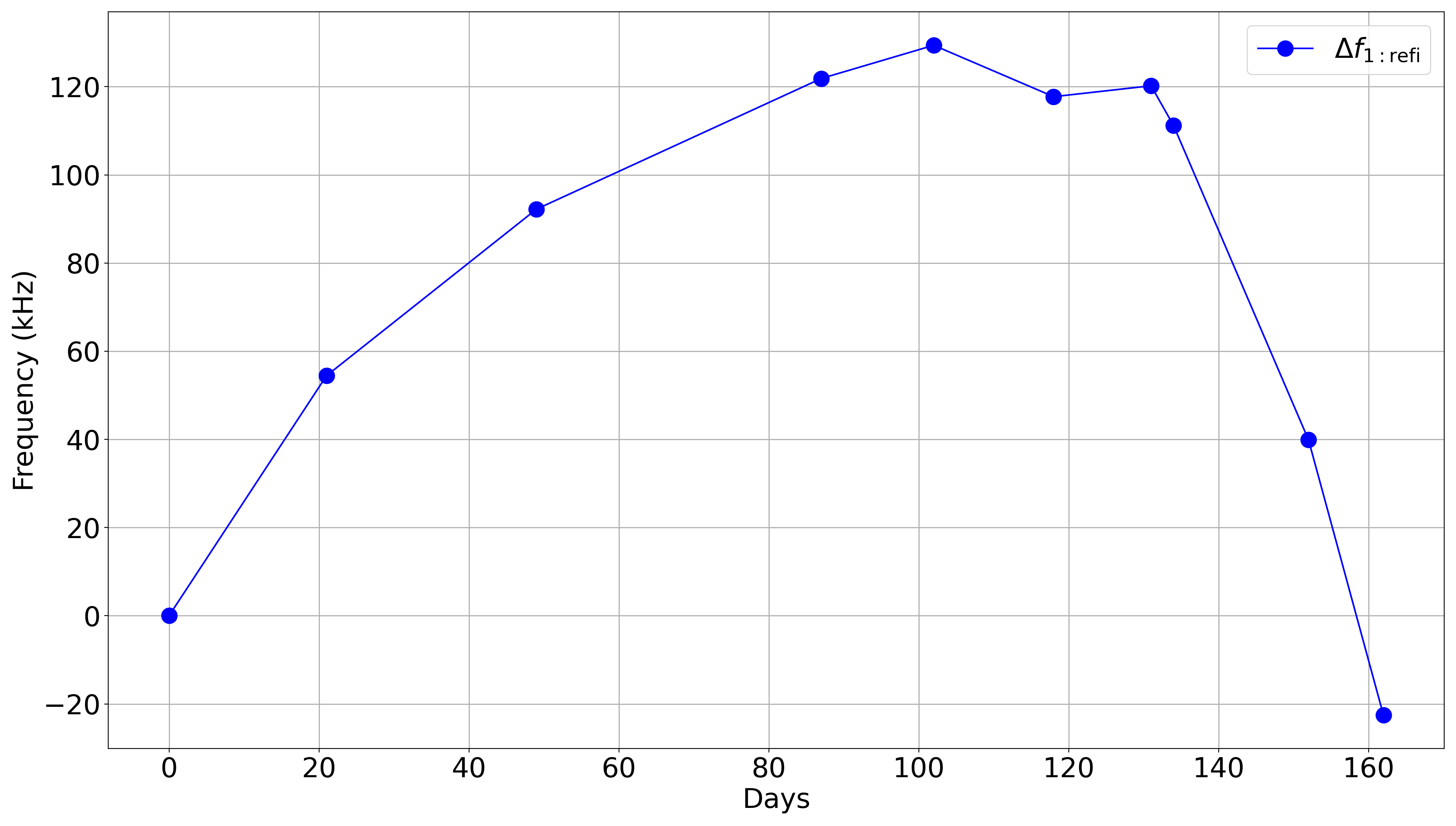}
\captionsetup{width=0.9\textwidth}
\caption{Measurement of the the long term drift of the LUS over the course of 160 days. Results shows the value of $\Delta f_{1:\text{refi}}= f_{1:\text{refi}}(\text{day})-f_{1:\text{refi}}(0)$}
\label{fig:drift_ULE}
\end{figure}

We have shown in the previous sections how the Refimeve signal is a powerful tool to carry out frequency stability or phase noise measurements of a laser under test, in our case the 729 nm local ultrastable laser used to stabilize the frequency comb.\\Another key feature enabled by this optical reference is the enhancement of the laser frequency determination that we can carry out. While the repetition frequency $f_{\text{rep}}$ of the OFC is locked to the LUS with the method previously described (see fig. \ref{fig:setup}), we can measure its frequency $\nu_{\text{LUS}}$ by measuring $f_{\text{rep}}$ and using the comb's formula :  
\begin{equation}
\nu_{\text{LUS}} = N_{\text{LUS}} \times f_{\text{rep}} \pm f_{\text{RF}}
\label{eqn:comb1}
\end{equation}
with $N_{\text{LUS}}$ the comb's mode number used to generate the beatnote $f_{\text{beat}}$ with the LUS. The locking scheme imposes \mbox{ $f_{\text{beat}}= f_{\text{RF}}$} with $f_\text{RF}$ the synthesizer carrier frequency used to create the error signal.
In this case the uncertainty $\delta \nu_{\text{LUS}}$ on the determination of the laser's frequency is proportional to the uncertainty $\delta f_{\text{rep}}$ on $f_{\text{rep}}$, with a scaling factor $N_{\text{LUS}}\approx$5$\times 10^{6}$. With adapted filtering and a 200 ms gate time on the phase counter, we typically reach $\delta f_{\text{rep}}$= 0.3 mHz in a 2 seconds averaging window, which leads to $\delta \nu_{\text{LUS}} \approx 1.5$ kHz.\\ Using the Refimeve signal with a known absolute frequency $\nu_{\text{refi}}= 194 400 008 500 000$~Hz~$\pm 5$Hz we can avoid measuring $f_{\text{rep}}$ from a beat between any adjacent comb's mode and rescale it to the optical domain, as it is usually done when no optical reference is available. Instead we significantly reduce $\delta \nu_{\text{LUS}}$ by expressing $f_{\text{rep}}$ as a function of $\nu_{\text{refi}}$
\begin{equation}
f_{\text{rep}} = \frac{\nu_{\text{refi}} \mp f_\text{1:refi}}{N_{\text{refi}}}
\label{eqn:comb2}
\end{equation}
with $N_{\text{refi}}$ the comb's mode number used to generate the beatnote $f_{\text{1:refi}}$. Re-writing Eq. (\ref{eqn:comb1}) as:
\begin{equation}
\nu_{\text{LUS}} = \frac{N_{\text{LUS}}}{N_{\text{refi}}}(\nu_{\text{refi}} \mp f_{\text{1:refi}}) \pm f_{RF}
\label{eqn:comb3}
\end{equation}
allows for a direct link between $\nu_{\text{LUS}}$ and $\nu_{\text{refi}}$ with a scaling factor $\frac{N_{\text{LUS}}}{N_{\text{refi}}} \approx 2.11$. With a typical  uncertainty $\delta f_{1:\text{refi}} \approx 5$Hz we reach $\delta \nu_{\text{LUS}} < 15$ Hz.\\ This represents a 100 fold improvement compared to the method previously used without the Refimeve signal. Note that we have neglected the uncertainty on $f_{\text{RF}}$ as it comes from a GPS-referenced low-noise synthesizer and is not impacted by the comb's scaling factor.\\
This gain on the frequency determination is propagated to any laser locked to the OFC and is also strongly enhancing the drift measurement of the LUS. Indeed a typical drift in the Hz$\cdot s^{-1}$ regime would be scaled down to the $\mu$Hz$\cdot s^{-1}$ level when measured on $f_{\text{rep}}$. This makes its measurement via $f_{\text{rep}}$ counting rather long due to $\delta f_{\text{rep}}$ and 30 min of acquisition were typically needed to measure a linear drift. Now using Refimeve, only 30 to 60 seconds are needed as the measured drift is only reduced by a factor 2.11. This allows us to design an active drift correction of the LUS, shifting the frequency of AOM$_1$ depending on a drift value updated every minutes.\\
Both these improvements represent a step towards a more precise and more accurate coherent spectroscopy on Ca$^+$ \cite{collombon_experimental_2019}. As an example, fig. \ref{fig:drift_ULE} shows the drift measurement we are carrying using the Refimeve signal to characterize the long term drift of the LUS, and therefore of ULE$_1$. With the enhanced resolution provided by the signal we are able to measure quicker, hence more often, and we can identify problems in our local frequency chain. Indeed we see on fig. \ref{fig:drift_ULE} at day 102 that the LUS step out of its natural "breathing" pattern due to a temperature stabilization failure of ULE$_1$.

\clearpage
\section{Conclusion}
The deployment of national and international compensated fibre links has already shown extensive use in various experimental physics fields. In France and Europe the Refimeve project is actively maintaining and developing a national network of such state-of-the art links, enabling the dissemination of an absolute, optical frequency reference at 1542 nm to about thirty users through more than 9000 km of fibres. Despite its intensive use there is, to our knowledge, no detailed analysis of the signal on a long-haul link. Being part of the network, and equipped with a LUS laser, we have carried out such an analysis by comparing the 1542 nm source laser to our local LUS at 729 nm through a frequency comb. We have detailed the experimental setup as well as the methodology put in place to compare both lasers and conclude on timescales of interest for the use of the Refimeve signal. Based on frequency instability measurement we've identified a short term region below 1 s  where the Refimeve signal stability is limited by the link. For longer measurement time the link noise is fully corrected and we can retrieve the plain stability of the Refimeve source laser, that we use to measure and correct the drift of the local LUS. By doing so we can identify a flicker frequency plateau at 2 $\times 10^{-15}$ on our system between 1 s and 12 s while for longer averaging time we find exact agreement with the source data. We overcome the difficulties of the local diagnosis and show how this metrological signal can be of use to measure laser drift and stability down to about 6 $\times 10^{-16}$ at 2 s and anyway below 5 $\times 10^{-15}$. We pursue in the same philosophy with a phase noise analysis of the Refimeve signal against the LUS. Similar conclusions are found and the observation window is extended to shorter timescale (higher Fourier frequencies), showing that, with the level of accuracy allowed by our set-up, the loop-back phase noise measurements on the Refimeve network is in perfect agreement with a end-user diagnostic. A laser linewidth is calculated from the phase noise measurement and shows that within 40 ms, the Refimeve signal reaches a stationary linewidth of 6.2 kHz, making it a tool of choice for any lab which is not equipped with NMI's grade laser systems.\\

\section*{Funding}
This work received support from the Région Sud under its exploratory research funding and the french government under the France 2030 investment plan, as part of the Initiative d'Excellence d'Aix-Marseille Université - AMIDEX AMX-23-REC-COFIRE-CG-02.\\ This work was also supported by the Labex Cluster of Excellence FIRST-TF (ANR-10-LABX-48-01), within the Program “Investissements d'Avenir” operated by the French National Research Agency (ANR), and by Equipex REFIMEVE+ (ANR-11-EQPX-0039) and by ESR/Equipex+ T-REFIMEVE (ANR-21-ESRE-0029). 
\section*{Acknowledgments}
We acknowledge the unwavering support of the network and engineering team of RENATER, especially Nicolas Quintin and Laurent Gydé.\\ M.C thanks Anne Amy-Klein and Rodolphe Le Targat for very helpful discussions about low noise frequency synthesis and transfer, as well as the Labex FIRST-TF for financial support.


\printbibliography

\end{document}